\documentclass[prl,twocolumn,floatfix,superscriptaddress,citeautoscript]{revtex4-2}
\usepackage{graphicx}
\usepackage{xcolor}
\usepackage{lineno}

\definecolor{mygray}{RGB}{100,100,100}

\usepackage{dcolumn}   
\usepackage{bm}        
\usepackage{amssymb}   
\usepackage{verbatim}
\usepackage{indentfirst}
\usepackage{multirow}
\usepackage{xcolor} 
\usepackage{amsmath}
\usepackage{amsfonts}
\usepackage{hyperref}
\usepackage[normalem]{ulem}
\hypersetup{colorlinks=true, pdfstartview=FitV, linkcolor=blue, citecolor=black, plainpages=false, pdfpagelabels=true, urlcolor=blue}
\usepackage[all]{hypcap}  

\newcommand{\state}[1]{|#1\rangle}
\newcommand{\statedagger}[1]{\langle#1|}
\newcommand{\expectation}[1]{\langle#1\rangle}

\newcommand{\pgithirteen}[0]{\affiliation{Institute for Functional Quantum System (PGI-13), Forschungszentrum J\"ulich, 52425 J\"ulich, Germany}}
\newcommand{\pgieight}[0]{\affiliation{Institute for Quantum Control (PGI-8), Forschungszentrum J\"ulich, 52425 J\"ulich, Germany}}
\newcommand{\rwth}[0]{\affiliation{Department of Physics, RWTH Aachen University, 52074 Aachen, Germany}}
\newcommand{\unikoln}[0]{
\affiliation{Institute for Theoretical Physics, University of Cologne, 50937 Cologne, Germany}}

\begin{document}
\title{Ultrafast Single Qubit Gates through Multi-Photon Transition Removal}
\author{Y. Gao}
\email[]{y.gao@fz-juelich.de}
\pgithirteen
\rwth

\author{A. Galicia}
\pgithirteen
\rwth

\author{J. D. Da Costa Jesus}
\pgieight
\unikoln

\author{Y. Liu}
\pgithirteen

\author{Y. Haddad}
\pgithirteen
\rwth

\author{D.~A.~Volkov}
\pgithirteen
\rwth

\author{J. R. Guimar\~aes}
\pgithirteen
\rwth

\author{H.~Bhardwaj}
\pgithirteen
\rwth

\author{M. Jerger}
\pgithirteen

\author{M. Neis}
\pgithirteen
\rwth

\author{B.~Li}
\pgieight

\author{F.~A.~C\'ardenas-L\'opez}
\pgieight

\author{F. Motzoi}
\pgieight
\unikoln

\author{P. A. Bushev}
\pgithirteen

\author{R. Barends}
\email[]{r.barends@fz-juelich.de}
\pgithirteen
\rwth

\date{\today}

\maketitle
\textbf{One of the main enablers in quantum computing is having qubit control that is precise and fast \cite{preskill_review,Surface_code_review,review_2025,roadmap}.
However, qubits typically have multilevel structures \cite{fluxonium,zero-pi,Transmon,flux_qubit} making them prone to unwanted transitions from fast gates. This leakage out of the computational subspace is especially detrimental to algorithms as it has been observed to cause long-lived errors, such as in quantum error correction \cite{LDPC,Matt_reset_QEC,QEC_threshold}. This forces a choice between either achieving fast gates or having low leakage. Previous works focus on suppressing leakage by mitigating the first to second excited state transition, overlooking multi-photon transitions, and achieving faster gates with further reductions in leakage has remained elusive. Here, we demonstrate single qubit gates with a total leakage error consistently below $2.0\times10^{-5}$, and obtain fidelities above $99.98\%$ for pulse durations down to 6.8~ns for both $X$ and $X/2$ gates. This is achieved by removing direct transitions beyond nearest-neighbor levels using a double recursive implementation of the Derivative Removal by Adiabatic Gate (DRAG) method, which we name the R2D method. 
Moreover, we find that  at such short gate durations and strong driving strengths the main error source is from these higher order transitions. 
This is all shown in the widely-used superconducting transmon qubit \cite{Transmon,Xmon}, which has a weakly anharmonic level structure and suffers from higher order transitions significantly. We also introduce an approach for amplifying leakage error that can precisely quantify leakage rates below $10^{-6}$. The presented approach can be readily applied to other qubit types as well.}

Superconducting qubits are engineered quantum devices that have multiple energy levels, distinct from natural two-level systems. Therefore, overcoming undesired transitions has been a constant challenge in the quantum control of such devices \cite{review_2005,DRAG_theory,Impulse_SQG,review_2019,digital_SQG}. The first works to mitigate this leakage used quadrature modulation to direct the quantum evolution, using the Derivative Removal by Adiabatic Gate (DRAG) method \cite{DRAG_theory}, which has since become standard practice. Later works have explored methods ranging from frequency domain optimization to piecewise constant pulse engineering \cite{IQM_FASTDRAG,DPSS_control,Active_cancelling_google, optimal_control_piecewise_constant}.
What has seemingly gone unrecognized, is that when pushing for performance with ever faster gates,
the qubit enters the strongly driven regime
where the driving strength approaches the anharmonicity, altering gate physics.

Higher order transitions beyond nearest-neighbor energy levels are normally either considered forbidden or can be safely neglected for weak driving. Yet, further leakage reduction is only possible by addressing these transitions, which appear during the evolution and can be revealed through the Schrieffer-Wolff (S-W) transform~\cite{REDRAG_first}. We start by describing the system in the rotating frame at the microwave driven frequency $\omega_d$. The Hamiltonian of a capacitively driven qubit can be written as~\cite{DRAG_theory},
\begin{equation}
    H/\hbar = \sum_j\delta_j\state{j}\statedagger{j}+\lambda_{j+1}\left(\frac{\Omega_R}{2}\sigma^x_{j,j+1}+\frac{\Omega_I}{2}\sigma^y_{j,j+1}\right)
    ,
\end{equation}
where $\delta_j = \omega_j-j\omega_d$ is the detuning of the $j$th energy level. $\Omega = \Omega_R+i\Omega_I$ is the control pulse, $\sigma^x_{j,k} = \state{j}\statedagger{k}+\state{k}\statedagger{j}$ and $\sigma^y_{j,k} = i\state{k}\statedagger{j}-i\state{j}\statedagger{k}$ are the Pauli operators, and $\lambda_j$ is the relative strength of different transitions, with $\lambda_1 = 1$. When the system is driven at resonance: $\omega_d=\omega_q=\omega_1$, and $\delta_2=\omega_2-2\omega_1=\Delta$ is the anharmonicity. In this qubit frame, the leakage from $\state{1}$ to $\state{2}$ can be straightforwardly identified through the $\sigma_{1,2}^x$ term. 
However, during the qubit driving, the qubit eigenstates are modified by the driving field, which leads to a superposition of higher qubit states, as well as shifts in eigenfrequencies. 
Therefore, to fully minimize the leakage out of the computational subspace with faster and stronger drives, we have to consider higher leakage levels and look at the dressed states of the coupled system formed by the qubit and the driving field.

\begin{figure}
    \includegraphics{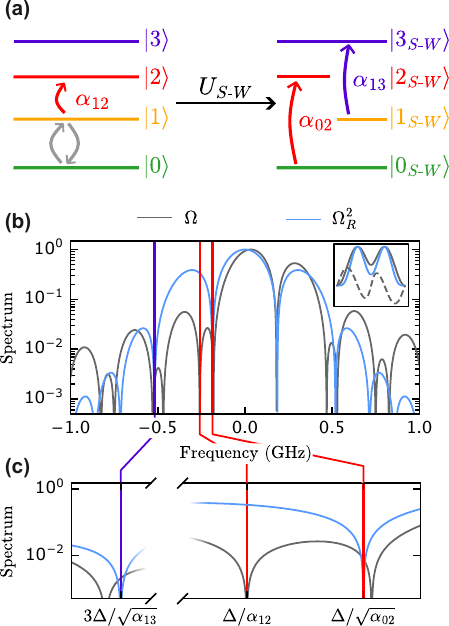}
    \caption{\textbf{Schematic of qubit level structures and pulse shapes for fast gates.} (a) Schematic representation of qubit energy levels in the qubit eigenbasis (left) and in the dressed eigenbasis under Schrieffer-Wolff transformation (right). The colored arrows show the unwanted leakage transitions. (b) An example R2D pulse for $t_p=7$~ns. The magnitude spectrum of both $\Omega = \Omega_R+i\Omega_I$ (gray) and $\Omega_R^2$ (blue) are plotted. The solid lines highlight the unwanted transition frequencies to $\state{2}$ (red) and $\state{3}$ (purple). The inset shows the normalized time domain pulse for both $\Omega$ and $\Omega_R^2$, and the dashed line shows $\Omega_I$. (c) The same magnitude spectrum of $\Omega$ and $\Omega_R^2$ as in (b), highlighting the frequencies of the unwanted single- and multi-photon transitions.}
    \label{fig:figure1}
\end{figure}

We now use the S-W transform on the system Hamiltonian. Using
$U_{S\text{-}W} = \exp(-i\Omega_R\sum_{j=0}^2\lambda_{j+1}\sigma^y_{j,j+1}/2\Delta)$, and keeping computationally relevant terms to the second order in $\Omega_R/\Delta$ we get~\cite{Jose}

\begin{equation}
\begin{split}
    H_{S\text{-}W}/\hbar  \approx \frac{\Omega_R}{2}  \sigma_{0,1}^x+ \frac{\Omega_I+\frac{\dot{\Omega}_R}{\Delta} }{2} \left(\sigma_{0,1}^y+\lambda_{2}\sigma_{1,2}^y\right)+ \\
         \left(\delta_1 +\frac{4-\lambda_{2}^2 }{4\Delta}\Omega_R^2 \right)|1\rangle\langle 1|   + \frac{\lambda_{2}}{8\Delta}\Omega_R^2 \sigma_{0,2}^x+\frac{\lambda_{2}\lambda_{3} }{8\Delta}\Omega_R^2 \sigma_{1,3}^x
         ,
\end{split}
\label{eq:H_SW}
\end{equation}
where the $\sigma_{0,2}^x$ and $\sigma_{1,3}^x$ terms in Eq.~(\ref{eq:H_SW}) reveal the direct transitions between higher levels. See Fig.~\ref{fig:figure1}(a) for a schematic representation. Interestingly, those transitions violate selection rules in the qubit frame. The two-photon nature of the transitions is further underlined by the squared term,~$\Omega_R^2$. 
Moreover, the quadratic dependence of the amplitude emphasizes its importance when the pulse duration is short.

With higher order transitions identified, we propose a pulse shape that suppresses the frequency spectrum of $\Omega_R^2$ at the relevant two-photon transition frequencies. Given a control envelope ansatz $\Omega_0$, we recursively apply the DRAG technique to remove the multi-photon transitions. To keep $\Omega_R^2$ real and avoid phase errors, we use the second derivative to remove the frequency components. A detailed proof and modeling are given in Ref.~\cite{Jose}. We also include parameters $\alpha_{02}$ and $\alpha_{13}$ which allow for fine-tuning the spectral notches during in-situ optimization,
\begin{equation}
\Omega_R^2 = \left(1+\frac{\alpha_{02}}{\Delta^2}\frac{d^2}{dt^2}\right)\left(1+\frac{\alpha_{13}}{9\Delta^2}\frac{d^2}{dt^2}\right)\Omega_0^2
.
\label{eq:squared_pulse}
\end{equation}
Subsequently, we apply the standard single-photon DRAG correction and include $\alpha_{12}$,
\begin{equation}
\Omega = \left(1-i\frac{\alpha_{12}}{\Delta}\frac{d}{dt}\right)\Omega_R
.
\label{eq:envelope}
\end{equation}
Similar to Ref.~\cite{DRAG_google}, we include a constant detuning to correct the phase error. 
To ensure that the derivatives exist and the pulse spectrum is finite, the envelope ansatz $\Omega_0$ must be smooth to at least 3rd-order and $\Omega(0) = \Omega(t_p) = 0$, where $t_p$ is the pulse length. In this work, we use $\Omega_0=A\sin^4(\pi t/t_p)$ and we add 1~ns padding after the pulse to account for possible pulse distortions in the system~\cite{distortion_oliver}. Using the Fourier transform, the pulse correction results in,
\begin{equation}
\left[\Omega_R^2\right]\left(\omega\right) = \left(1-\frac{\alpha_{02}\omega^2}{\Delta^2}\right)\left(1-\frac{\alpha_{13}\omega^2}{9\Delta^2}\right)\left[\Omega_0^2\right]\left(\omega\right)
,
\end{equation}
\begin{equation}
\Omega\left(\omega\right) = \left(1-\frac{\alpha_{12}\omega}{\Delta}\right)\Omega_R\left(\omega\right)
.
\end{equation}
The effect of the R2D pulse shaping can be intuitively understood in the frequency domain of both $\Omega$ and $\Omega^2_R$~[Fig.~\ref{fig:figure1}(b)-(c)], where we null the signal spectrum at leakage transition frequencies.

We experimentally validate the R2D pulse shape using an isolated transmon qubit  (see Supplement). The qubit is operated at the flux insensitive spot where it has a frequency $f_q$ of 5.323~GHz and an anharmonicity $\Delta/2\pi$ of -225~MHz.

We start the validation by introducing a method capable of measuring a leakage error down to $10^{-6}$.
The standard approach uses leakage randomized benchmarking (LRB)~\cite{LRB_theory}, where the leakage is measured as it accumulates during Clifford-based randomized benchmarking. However, this approach has drawbacks. First, the random nature causes the accumulation of higher level state population to be incoherent. Second, extracting the leakage error of a specific gate requires interleaving, which can become expensive in terms of time and measurements. Third, since LRB accumulates leakage incoherently, it becomes challenging to distinguish coherent gate leakages from thermal transitions. Moreover, we show later that it becomes difficult to obtain precision below $10^{-5}$ using LRB. Therefore, we introduce an amplified leakage error (ALE) method that coherently accumulates leakage errors for both $X$ and $X/2$ gates, allowing for highly accurate measurements.

\begin{figure}
    \includegraphics{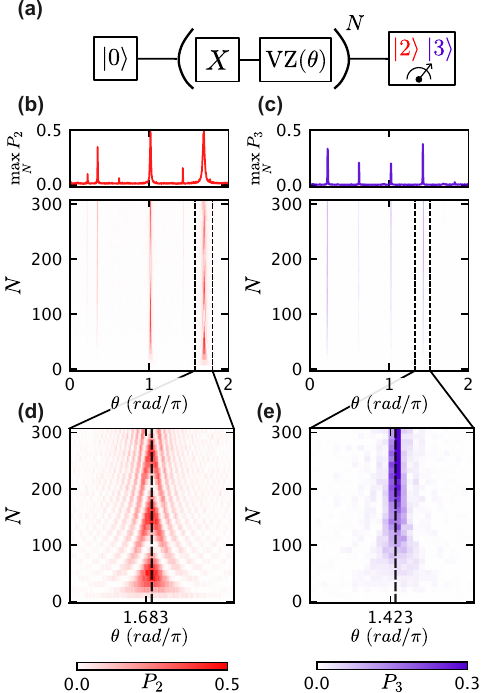}
    \caption{\textbf{Amplified leakage error experiments.} (a) The schematic of the ALE pulse sequence. We repeatedly apply $N$ times a $X$ operation followed by a VZ with angle $\theta$. The sequence ends with measurements of the leakage population. (b) The $\state{2}$ leakage population as a function of $\theta$ and $N$. The line plot on top shows the maximum leakage population for each $\theta$. (c) Same as (b) but plotted with $\state{3}$ leakage population. (d)(e) The same data as in (b)(c) but with a magnified window around one of the amplification phases (dashed lines). The clear chevron pattern indicates a coherent oscillation of the leakage population.}
    \label{fig:ale_intro}
\end{figure}
The key notion behind the ALE sequence is the phase matching of the leakage population after each operation. Using a $Z$-rotation with correct angle, the population will be phase matched and coherently accumulated. The ALE sequence consists of $N$ repetitions of an $X$ or composite $X$ operation and a virtual-$Z$ (VZ) gate with angle~$\theta$~[Fig.~\ref{fig:ale_intro}(a)]. A similar principle was previously employed to amplify errors in two-qubit gates~\cite{2Q_ALE,BL_CR}. The unitary of $X$ containing coherent leakages and the VZ can be written as
\begin{align}
X &= \begin{pmatrix}
0&1&A_1e^{-i(\phi_1+\phi_2)}\\
1&0&A_0e^{-i(\phi_0+\phi_2)}\\
-A_0e^{i(\phi_0-\phi_2)}&-A_1e^{i(\phi_1-\phi_0)}&e^{-i2\phi_2}
\end{pmatrix}
, \\
\textrm{VZ}(\theta) &= \begin{pmatrix}1&0&0\\0&e^{-i\theta}&0\\0&0&e^{-i2\theta}\end{pmatrix}
,
\end{align}
where $A_0,~A_1\ll1$ are magnitudes of the leakage population transferred to $\state{2}$, $\phi_0$ and $\phi_1$ are phases of the leakage population, and $2\phi_2$ is the phase acquired in $\state{2}$ during $X$. The leakage rate of $X$ is defined as $\epsilon_{leak,\state{2}}=(A_0^2+A_1^2)/2$. After we examine the ALE sequence, we find (see Supplement) that leakage is coherently accumulated only when $\theta$ satisfies
\begin{equation}
3\theta+4\phi_2=2k\pi,~k\in\mathbb{Z}
.
\label{eq:target_phase}
\end{equation}
At the phase $\theta_k = (2k\pi-4\phi_2)/3$, the qubit population coherently oscillates between the computational subspace and $\state{2}$, and~$P_2 = a_k\sin^2(\omega_kN/2)$, where $a_k$ and $\omega_k$ are the amplitude and the frequency of the oscillation. The leakage error of $X$ is proportional to the weighted sum of squared oscillation frequencies at all amplification phases, 
\begin{equation}
    \epsilon_{leak,\state{2}} = \frac{1}{6}\sum_{k=1}^3a_k\omega_k^2.
\end{equation} 
Similar analysis can be applied for the leakage rate to $\state{3}$, where five distinct phases exist (see Supplement). While the coherent leakage error caused by the control pulse is amplified, the thermal leakage with random phases will remain low. Therefore, the ALE sequence can selectively amplify the coherent leakage error to different states.

\begin{figure*}[!t]
    \includegraphics{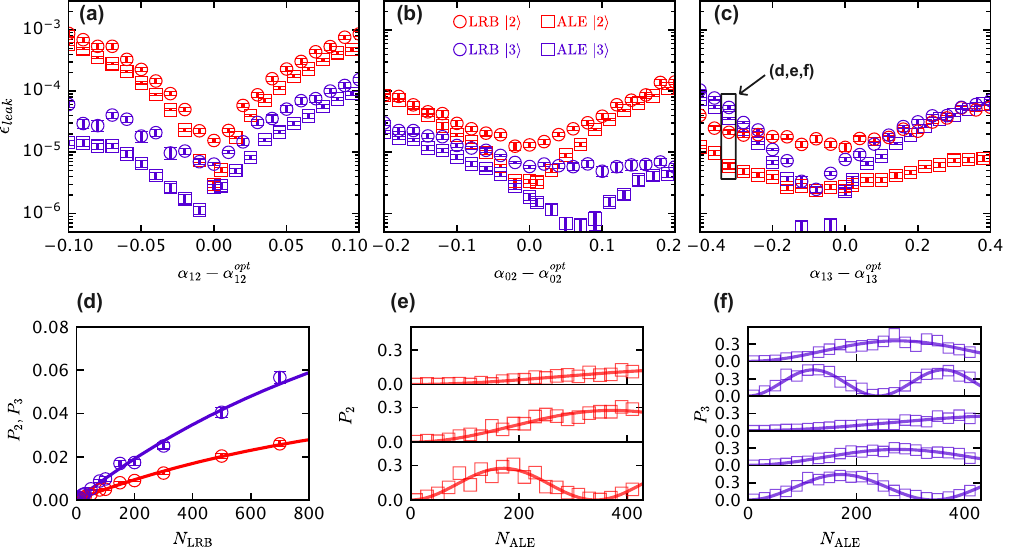}
    \caption{\textbf{Extracting leakage error from leakage randomized benchmarking and amplified leakage error.} (a)(b)(c)~Leakage error rates of an X gate at $t_p=7$~ns extracted from LRB (circle) and ALE (square) for leakage state $\state{2}$ (red) and $\state{3}$ (purple) as a function of parameter $\alpha_{12}$ (a), $\alpha_{02}$ (b), and $\alpha_{13}$ (c). The optimal parameter set $(\alpha_{12}^{opt},\alpha_{02}^{opt},\alpha_{13}^{opt})$ is obtained through an in-situ optimization. The raw traces of the marked data points in (c) are shown. (d) Raw traces used to extract leakage errors marked in (c) for both states using LRB. Only the interleaved curves are shown. Solid lines are fit curves. (e)(f) Raw traces used to extract leakage errors marked in (c) for $\state{2}$ (e) and $\state{3}$ (f) using ALE. Solid lines are fit curves.}
    \label{fig:parameter_sweep}
\end{figure*}

We show an example of the experimental implementation of the ALE sequence in Fig.~\ref{fig:ale_intro}(b)-(e). The leakage population is plotted as a function of $N$ and $\theta$. The amplified leakage error to $\state{2}$ is on the left, and error to $\state{3}$ is on the right. Despite the existence of state preparation and measurement (SPAM) errors, the amplification of leakages to different states is independent and clearly visible. The line plots in Fig.~\ref{fig:ale_intro}(b)-(c) show the maximum leakage population over $N$ as a function of $\theta$. The target phases which amplify the leakage error are evenly spaced in $[0,2\pi)$. The small peaks in Fig.~\ref{fig:ale_intro}(b) are caused by SPAM errors. Finally, in Fig.~\ref{fig:ale_intro}(d)-(e), we see clear chevron patterns evidencing a coherent process, with dashed lines at target phases.

The ALE experiment used in the gate calibration procedure can be further simplified. In fact, the target phases satisfying Eq.~(\ref{eq:target_phase}) stay unchanged unless $\phi_2$ varies. As $\phi_2$ is dominated by the frame frequency of $\state{2}$, the target phases are stable in a large range of parameters for a constant pulse duration. Thus, only a $2\pi/3$ window is needed to find all phases, and a narrow window around them can be used afterwards. This significantly reduces the calibration time needed in the close-loop optimization. 

The amplified leakage error sequence complements the metrology and calibration toolset for characterizing all errors of single qubit gates, together with the amplified phase error (APE) sequence~\cite{APE} and the amplified amplitude error (AAE) sequence~\cite{AAE}. For each set of parameters $(\alpha_{12},\alpha_{02},\alpha_{13})$, APE and AAE experiments can deterministically find out the optimal constant detuning and pulse amplitude (see Supplement). Using the leakage error extracted from ALE experiments as the cost function, the Nelder-Mead optimization can reliably find the optimal gate under the R2D pulse shaping. 

We validate the leakage errors obtained from the ALE through a comparison with the standard LRB method. 
In Fig.~\ref{fig:parameter_sweep}(a)-(c), leakage errors of an $X$ gate at $t_p=7$~ns from both methods are plotted as a function of $\alpha_{12}$, $\alpha_{02}$, and $\alpha_{13}$, respectively. The parameters are individually swept around the optimal value obtained from an in-situ gate calibration, $(\alpha_{12}^{opt},\alpha_{02}^{opt},\alpha_{13}^{opt})=(0.864,1.467,1.796)$. For each parameter set, APE and AAE experiments are used to find the minimum amplitude and phase error. 
A set of raw traces for the data points marked in Fig.~\ref{fig:parameter_sweep}(c) from both methods are shown in Fig.~\ref{fig:parameter_sweep}(d)-(f). The incoherently accumulated leakage of interleaved LRB, averaged over 60 random sequences, as well as the solid fit lines are plotted in Fig.~\ref{fig:parameter_sweep}(d). 
The population stays below~0.06 even after 800 Clifford gates. On the other hand, the amplified leakages from ALE experiments for $\state{2}$ and $\state{3}$ show clear coherent oscillations at amplification phases within just 400 operations, see the fitted curves in Fig.~\ref{fig:parameter_sweep}(e)-(f). 
As shown in Fig.~\ref{fig:parameter_sweep}(a)-(c), we find that the leakage rates increase by almost two orders of magnitude when the parameters are biased away from the optimal value. 
While a good correspondence of the leakage rates between two methods is prominently visible, noticeable differences appear for leakages below $10^{-5}$. 
Despite differences in leakage errors between methods, the minimum total leakage is achieved with the same pulse parameters.

The leakage behavior in Fig.~\ref{fig:parameter_sweep} provides an intuitive guidance for optimizing pulse parameters. After minimizing the phase and amplitude error, the leakage is solely controlled by $(\alpha_{12},\alpha_{02},\alpha_{13})$, and the optimal value can be obtained through the Nelder-Mead optimization. The corrections from $\alpha_{02}$ and $\alpha_{13}$ are nearly orthogonal, which benefits the optimization. 
Additionally, the ALE sequence requires less operations and shows larger leakage populations, which emphasizes its advantages in terms of time and measurements. 
Furthermore, a reliable gate optimization can be achieved through ALE instead of LRB, evidenced by the good correspondence in both leakage errors and the optimal parameter values. We attribute the differences in leakage errors to the fact that the ALE method is insensitive to thermal excitations and $\state{2}\rightarrow\state{3}$ transition errors.

In order to evaluate the gate performance from all aspects, we need additional benchmarking tools.
Besides LRB characterizing the leakage error $\epsilon_{leak}$, we use standard randomized benchmarking (RB) \cite{RB} to characterize the total gate error $\epsilon_{tot}$, and purity randomized benchmarking (PRB) \cite{PRB} to characterize the decoherence error $\epsilon_{decoh}$ (see Supplement). All methods are run in an interleaved fashion to extract errors for the target gate. 

\begin{figure}
    \includegraphics[width=1\linewidth]{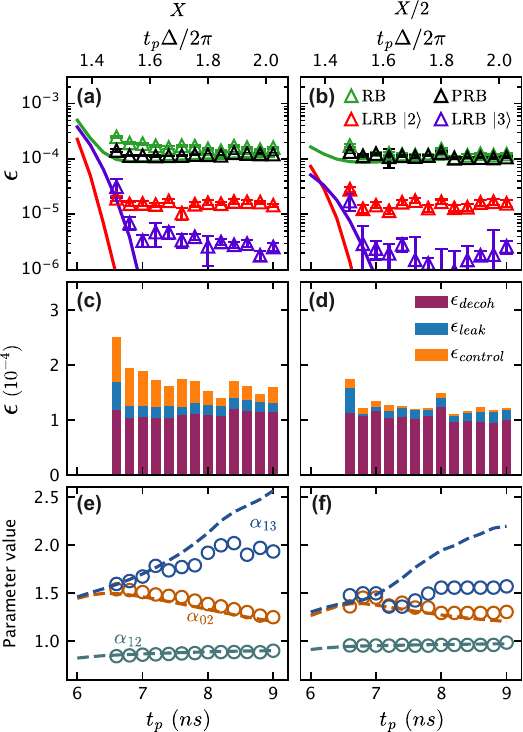}
    \caption{\textbf{Gate benchmarks with varying pulse lengths.} Gate performance with pulse durations $t_p\in [6.6,9]$ ns. X gate is on the left, and X/2 gate is on the right. (a)(b) Error rates extracted from RB (green), PRB (black), LRB for $\state{2}$ (red), and LRB for $\state{3}$ (purple). Solid lines are simulation results. (c)(d) Different contributions to total gate error, including decoherence error (dark magenta), leakage error (blue), and coherent control error (orange). (e)(f) Parameter sets of optimal gates from in-situ calibrations (symbols) and simulations (dashed lines).}
    \label{fig:rb_scan}
\end{figure}

To fully demonstrate the performance of the R2D pulse shaping, we perform a complete benchmark of single qubit gates with varying pulse lengths, and construct the error budgets.  Using the calibration procedure discussed previously, we bring up both $X$ and $X/2$ gates with pulse length $t_p\in[6.6 , 9]$~ns. Pulses
shorter than 6.6 ns will result in an unstable gate optimization as negative values appear in $\Omega_R^2$, and pulses longer than 9 ns will cause benchmarking sequences to exceed the waveform memory limit of the control electronics. To ensure the best results, a constant phase correction is included in the pulse (see Supplement) to compensate for the accumulated phase shift caused by the constant detuning.
In Fig.~\ref{fig:rb_scan}, we demonstrate the benchmarking results of $X$ gates on the left, and $X/2$ gates on the right. Error rates from all benchmarks are shown in Fig.~\ref{fig:rb_scan}(a)-(b). We see that the total leakage error stays below $2\times10^{-5}$ down to a pulse length of 6.8 ns. At a pulse duration of 6.6 ns, the leakage error to $\state{3}$ starts to dominate.
At a pulse duration of 6.8 ns, we achieve $X$ and $X/2$ gates with a total gate error of $(1.9\pm0.1)\times10^{-4}$ and $(1.1\pm0.1)\times10^{-4}$ respectively. 
We note that, with $t_\pi$ being the $\pi$-pulse duration, this is equicalent to a normalized driven strength of $\Omega/\Delta = 1/2t_{\pi}\Delta = 0.33$, a record value for single qubit gates in superconducting qubits.
The solid green line shows the total gate error from a Lindblad master equation simulation.
The solid purple line shows the simulated coherent leakage error to $\state{3}$ in the absence of incoherent processes. The same simulated leakage error to $\state{2}$ is plotted in red. 
We use the experiment results to separate the total gate error into constituents: $\epsilon_{tot} = \epsilon_{decoh}+\epsilon_{leak}+\epsilon_{control}$, see Fig.~\ref{fig:rb_scan}(c)-(d).
We find that gate errors of both $X$ and $X/2$ are mainly limited by decoherence errors around a level of $1.04\times10^{-4}$, and coherent control errors increase with decreasing pulse lengths for $X$ gates. We also show the optimal parameters for different pulse lengths in Fig.~\ref{fig:rb_scan}(e)-(f), where symbols and dashed lines represent values from the in-situ calibration and the simulation respectively.

In addition, when comparing the R2D pulse method we find it outperforms the standard single-photon DRAG correction (see Supplement), achieving a leakage reduction exceeding a factor of 20.

The data in Fig.~\ref{fig:rb_scan} show that the R2D pulse shape enables gates with nearly $99.99\%$ fidelity 
at the normalized driven strength $\Omega/\Delta$ up to 0.33,
limited by qubit decoherence. First, the leakage errors remain below $2.0\times10^{-5}$ over a large domain, and the increase in leakage errors below a pulse length of 6.8~ns is well predicted by the simulated coherent leakage. This evidently demonstrates the importance of accounting for multi-photon leakages in single qubit gates. We attribute the lower bound of leakages to excitations dominated by the voltage noise of room temperature electronics for the microwave control. Second, both $X$ and $X/2$ gates show small coherent errors, with the error being substantially below $1.0\times10^{-4}$ for $X/2$ gates. For $X$ gates, we attribute the coherent error increase with decreasing duration to microwave distortions, as it becomes more noticeable for shorter gates and higher amplitudes. This can be mitigated by introducing a small padding after the pulse. Third, the optimal pulse parameters from simulation and in-situ calibration compare well. 
Parameter $\alpha_{13}$ differs more for longer pulses because the change in $\state{3}$ leakage is below the measurement threshold.
The simulated and experimental results being consistent in all aspects validates the R2D pulse shape.

In conclusion, we demonstrate fast and precise $X$ and $X/2$ gates through suppression of multi-photon transitions in the strongly driven regime using only three parameters. 
We also introduce the ALE sequence which allows precision near $10^{-6}$ in coherent leakage characterization, and complements the metrology toolset for the single qubit gate calibration. The presented approaches can be straightforwardly adopted by other platforms that use microwave qubit control. In addition, the ability to suppress multiple transitions is also beneficial to driving qubits simultaneously. Moreover, the same approaches can be used to mitigate the leakage in fast two-qubit gates. These results set a new benchmark for single qubit gates in terms of both fidelities and speed, and with enhanced qubit coherence and precise distortion characterization, the proposed methodology opens a path to achieving gates with a fidelity of $99.999\%$.

\section{ACKNOWLEDGMENTS}
The authors are grateful for TWPA calibration support from Marcello Guardascione, data visualization support from Niklas Oertel, software support from Juan Cereijo, and theory support from Dimitrios Georgiadis. The device is made in the Helmholtz Nano Facility. We acknowledge the support by the German Federal Ministry of Research, Technology and Space (BMFTR), funding program “Quantum technologies - from basic research to market”, project QSolid (Grant No. 13N16149). We acknowledge funding from the Horizon Europe
program (HORIZON-CL4-2021-DIGITAL-EMERGING-
02-10) via the project 101080085 (QCFD) and by
HORIZON-CL4-2022-QUANTUM-01-SGA Project under Grant 101113946 OpenSuperQPlus100.

\section{CONTRIBUTIONS}

Y.G. and R.B. designed the experiments. Y.G. performed the experiments and analyzed the data. Y.G. conducted the simulation. J.D.D.C.J. and F.M. provided theoretical framework and models. Y.G. and A.G. implemented key components of the data-taking infrastructure. Y.G. designed the device. Y.L., Y.H., D.A.V., J.R.G., H.B., and M.N. contributed to device fabrication. Y.G. and R.B. prepared the manuscript. All authors contributed to the experimental infrastructures or the theoretical studies, and to the manuscript revision.

\section{DATA AVAILABILITY}
The data that support the findings of this study are available from the corresponding author upon reasonable request.

\bibliography{reference}

\clearpage

\setcounter{figure}{0}
\setcounter{equation}{0}
\renewcommand{\thefigure}{S\arabic{figure}}
\renewcommand{\theequation}{S\arabic{equation}}
\begin{titlepage}   
\centering
{\large\bfseries Supplementary Materials for\\ ``Ultrafast Single Qubit Gates through Multi-Photon Transition Removal"\par}
\vspace{1em}
{\normalsize Y. Gao,$^{\text{1, 2}}$ A. Galicia,$^{\text{1, 2}}$ J. D. Da Costa Jesus,$^{\text{3, 4}}$ Y. Liu,$^{\text{1}}$ Y. Haddad,$^{\text{1, 2}}$\\D. A. Volkov,$^{\text{1, 2}}$ J. R. Guimar\~aes,$^{\text{1, 2}}$ H. Bhardwaj,$^{\text{1, 2}}$ M. Jerger,$^{\text{1}}$ M. Neis,$^{\text{1, 2}}$\\
B. Li,$^{\text{3}}$ F. A. C\'ardenas-L\'opez,$^{\text{3}}$ F. Motzoi,$^{\text{3, 4}}$ P. A. Bushev,$^{\text{1}}$ and R. Barends,$^{\text{1, 2}}$\\}
\vspace{0.5em}
{\small\textit{$^1$Institute for Functional Quantum System (PGI-13), Forschungszentrum J\"ulich, 52425 J\"ulich, Germany\\
$^2$Department of Physics, RWTH Aachen University, 52074 Aachen, Germany\\
$^3$Institute for Quantum Control (PGI-8), Forschungszentrum J\"ulich, 52425 J\"ulich, Germany\\
$^4$Institute for Theoretical Physics, University of Cologne, 50937 Cologne, Germany}}\\
{(Dated: \today)}
\vfill\vspace{2.5em}
\end{titlepage}
\section{QUBIT PARAMETERS AND STATE DISCRIMINATION}

In this work, we use an isolated tunable transmon qubit that does not couple to any other qubits or couplers. The qubit chip is fabricated using the E-beam lithography. The $\text{Al-Al}_2\text{O}_3\text{-Al}$ junctions are defined using the standard Manhattan process on a Si substrate. 

The qubit has $E_C/h=204.8$~MHz and $E_J/h=18.7$~GHz. The operating point is near the flux insensitive point, where the qubit is at the frequency $f_q = 5.323$~GHz and has anharmonicity $\Delta/2\pi = -225.0$~MHz. The qubit XY control line is capacitively coupled to the qubit, and the coupling capacitance is $C_c = 95$~aF. The qubit is dispersively readout from a quarter-wave resonator. The resonator is at the frequency $f_r=6.799$~GHz, and has a qubit-resonator coupling of $g_{qr}/2\pi=45$~MHz. The average qubit coherence properties are $T_1=40$~$\mu$s from $T_1$ experiments, $T_2^E = 56$~$\mu$s from Echo experiments, and $T_2^R=45$~$\mu$s from Ramsey experiments.

\begin{figure}[h]
\includegraphics[width=1\linewidth]{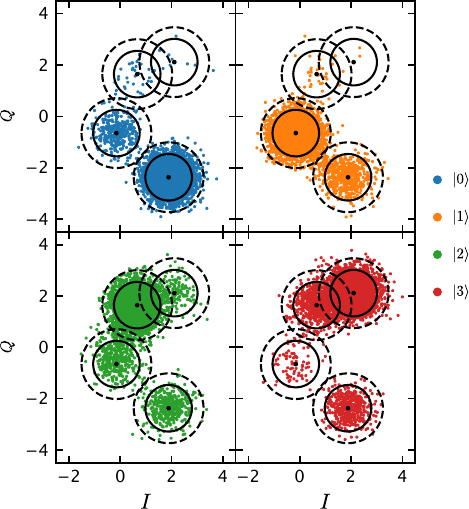}
\caption{ Raw IQ points from the single shot experiment where qubit is prepared in the $\state{0}$ (blue), $\state{1}$ (orange), $\state{2}$ (green), and $\state{3}$ (red) states. We show 5000 out of $2^{17}$ preparations for each state. The black points are the fitted state centers, the circles have the radii corresponding to two (solid line) and three (dashed line) standard deviations of the data.}
\label{fig:singleshot}
\end{figure}

To better quantify the leakage rate to different states, we use four-state discrimination. We reset the qubit to $\state{0}$ by relaxation, and measure $2^{17}=131072$ shots in each state. The acquisition length is 2.048~$\mu$s. The raw IQ points and the fitted 2D Gaussian distributions are shown in Fig.~\ref{fig:singleshot}. To discriminate the readout data, the nearest center method is used. We obtain the assignment matrix as
\begin{equation}
    \Lambda=\begin{pmatrix}
        89.64&9.49&0.75&0.12\\
        13.10&86.03&0.80&0.07\\
        13.67&11.43&69.93&4.97\\
        10.46&2.76&15.78&71.00
    \end{pmatrix}\times10^{-2},
\end{equation}
where $\Lambda_{ij}$, the element at $i$-th row and $j$-th column, indicates the probability of measuring $\state{j}$ when qubit is prepared at $\state{i}$. In this assignment matrix, a large number of thermal population $\bar n\approx0.1$ is observed. We use postselection to reduce the initialization error, and the new assignment matrix is 
\begin{equation}
    \Lambda^{ps}=\begin{pmatrix}
        97.793&2.129&0.073&0.007\\
        7.873&91.604&0.519&0.004\\
        6.726&11.173&77.602&4.499\\
        3.116&2.082&15.958&78.843
    \end{pmatrix}\times10^{-2}.
\end{equation}

In all experiments performed in this work, we use postselection to reset the qubit to $\state{0}$, and we do not correct for SPAM errors. 

\section{CONSTANT DETUNING AND PHASE CORRECTION}
\begin{figure}
\includegraphics[width=0.7\linewidth]{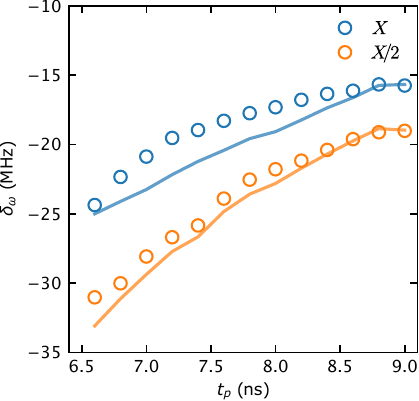}
\caption{Constant detuning corrections for different pulse durations from experiments (symbols) and predictions using the experimentally calibrated $\alpha_{12}$ (lines).}
\label{fig:const_detune_prediction}
\end{figure}
In this work, we use a constant detuning to correct the phase error. The theoretical value for the constant detuning $\delta_\omega$ satisfies 
\begin{equation}
    \delta_\omega = 0.712\frac{(4\alpha_{12}-2)}{\Delta}\frac{\pi^2}{t_g^2},
\end{equation}
see Ref.~\cite{Jose} for detailed derivations.
It matches well with experimental results [Fig.~\ref{fig:const_detune_prediction}], and can be used as initial values for the pulse. Additional to the constant detuning correction, an extra phase correction is included to compensate the accumulated phase shift, 
\begin{equation}
\Omega'(t) = \Omega(t) e^{-i\delta_\omega\left(t-t_p/2\right)}
,
\end{equation}
where $\Omega(t)$ is the pulse envelope as in Eq.~(4) in the main text. The phase correction $\delta_{\omega}t_p/2$ is particularly important when other single qubit gates of the same qubit are at different lengths, or their phase errors are corrected by other approaches. In the absence of this phase correction, the effective rotation axis will be away from $x$ axis in the $x-y$ plane. This explains the issue reported in Ref.~\cite{DRAG_google} that the best results are only achievable by using the same detuning and the same length for $X$ and $X/2$ gates. 

\section{SINGLE QUBIT GATE SIMULATION}

\begin{figure}[h]
\includegraphics[width=1\linewidth]{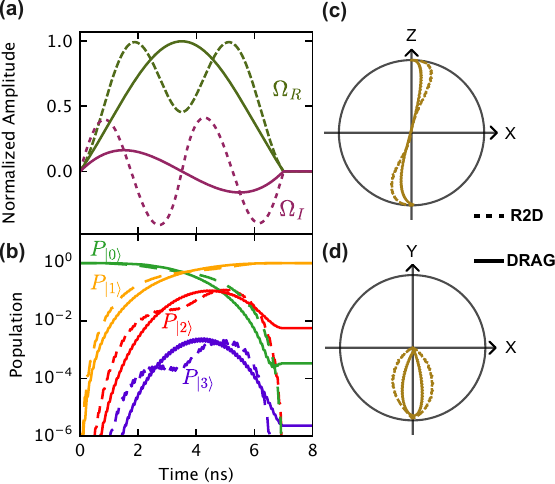}
\caption{ Simulation results of the optimal pulses using R2D (dashed lines) and DRAG (solid lines) pulse shapes. (a) The time domain signals of the real part and the imaginary part. (b) Simulated population of lowest four states. (c) Qubit state trajectories projected in $X\text{-}Z$ plane.  (d) Qubit state trajectories projected in $X\text{-}Y$ plane.}
\label{fig:simulation}
\end{figure}

We simulate the qubit evolution without RWA, as the rotating wave approximation (RWA) is only valid under weak coupling \cite{non-RWA}. The bare transmon is modeled in the qubit charge basis as 
\begin{equation}
H_q = 4E_C\hat{n}^2+E_J\cos\hat\phi,
\label{eq:transmon hamiltonian}
\end{equation}
where $\hat{n}$ is the charge operator, $\hat\phi$ is the phase operator, and 
\begin{equation}
\cos\hat\phi = \frac{1}{2}\sum_n\state{n}\statedagger{n+1}+\state{n+1}\statedagger{n}.
\end{equation}
Transmon parameters $E_C$ and $E_J$ are set such that the lowest four energy levels compare well with experiment results. There are in total 19 charge states in the simulation. A capacitively coupled microwave drive is modeled as
\begin{equation}
    H_d = \hbar\Omega_{RF}(t)\hat{n}.
\end{equation}
Microwave distortion is not considered. We also include 1 ns padding after the pulse as in the experiment. The pulse is modulated by the qubit frequency obtained from Eq.~(\ref{eq:transmon hamiltonian}). The QuTiP~\cite{Qutip} package is used to simulate the qubit evolution. In Fig.~\ref{fig:simulation}, simulation results of the optimal pulses at $t_p=7$~ns using R2D and DRAG pulse shapes are shown.

To incorporate incoherent processes during the evolution, we use Lindblad master equation,
\begin{equation}
    \dot\rho = \frac{i}{\hbar}[\rho,H_q+H_d]+\sum_iL_i\rho L_i^\dag-\frac{1}{2}\left\{\rho, L_i^\dag L_i\right\}
\end{equation}
where $\rho$ is the density operator, and we use collapse operators $L_- = \sqrt{\Gamma_\downarrow}a$, $L_+ = \sqrt{\Gamma_\uparrow}a^\dag$, and $L_\phi = \sqrt{\Gamma_\phi}a^\dag a$. $\Gamma_\downarrow = (1-\bar n)/T_1$, $\Gamma_\uparrow = \bar n/T_1$, and $\Gamma_\phi = 1/T_2^E-1/2T_1$ can be measured in experiments. 

\section{PULSE SHAPE IN TIME AND FREQUENCY DOMAIN}

\begin{figure}
\includegraphics[width=1\linewidth]{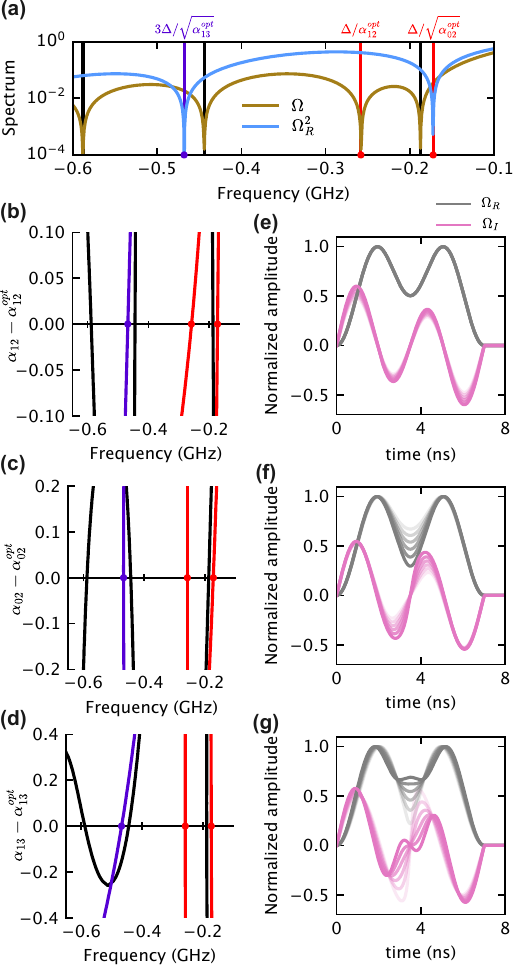}
\caption{ R2D pulse shape in the frequency domain and the time domain. (a) The optimal pulse at $t_p=7$~ns with pulse parameter $(\alpha_{12}^{opt},\alpha_{02}^{opt},\alpha_{13}^{opt})=(0.864,1.467,1.796)$. (b) The frequency domain pulses when parameter $\alpha_{12}$ is swept around $\alpha_{12}^{opt}$. The colored lines are the notches shown in (a), and the corresponding points are marked with dots. (e) The time domain pulses when parameter $\alpha_{12}$ is swept around $\alpha_{12}^{opt}$. The lighter color indicates a smaller value of the parameter. (c)(f) Same plots as in (b)(e) but sweeping $\alpha_{02}$. (d)(g) Same plots as in (b)(e) but sweeping $\alpha_{13}$. }
\label{fig:TF domain}
\end{figure}

It is important to understand in what way the corrections are affecting the pulse shape. In Fig.~\ref{fig:TF domain}, we show how the time and frequency domain pulses change when the pulse parameters are swept. In Fig.~\ref{fig:TF domain}(a), the spectrum of an optimal pulse at $t_p=7$~ns is plotted, and the parameters $(\alpha_{12}^{opt},\alpha_{02}^{opt},\alpha_{13}^{opt})=(0.864,1.467,1.796)$ are obtained from in-situ optimization. We mark the notches that we actively optimize with colors and the rest of notches black. The pulse parameters are swept individually around the optimal value and the pulses are plotted in both time and frequency domain. For time domain pulse, we use a lighter color to indicate a smaller value of the parameter. We sweep $\alpha_{12}$ in Fig.~\ref{fig:TF domain}(b)(e), $\alpha_{02}$ in Fig.~\ref{fig:TF domain}(c)(f), and $\alpha_{13}$ in Fig.~\ref{fig:TF domain}(d)(g).  From the frequency domain notches, the corrections from $\alpha_{02}$ and $\alpha_{13}$ are nearly orthogonal, which agrees with the leakage rates scan in Fig.~3 in the main text. Note the opposite effects on $\Omega_R(t)$ as $\alpha_{02}$ and $\alpha_{13}$ increase, this enables a short pulse since larger $\alpha_{02}$ corrections in fast pulses tends to make $\Omega_R^2$ negative.

\section{EXPERIMENTAL SETUP AND SIGNAL DISTORTION}

\begin{figure}
    \includegraphics[width=1\linewidth]{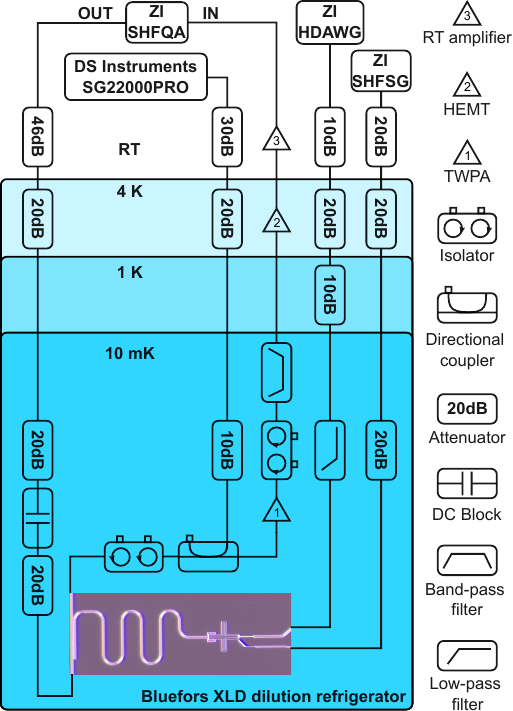}
    \caption{ The electronics and control wiring schematic. }
    \label{fig:wiring}
\end{figure}

The experimental setup and an optical picture of the transmon are shown in Fig.~\ref{fig:wiring}. 
The transmon qubit chip is packaged, and protected by a $\mu$-metal shield. The device is mounted at the mixing chamber of a Bluefors XLD fridge. The qubit flux is directly biased by a Z\"urich Instruments HDAWG using the voltage offset of the channel. A Z\"urich Instruments SHFQA is used for the qubit readout. The readout channel has $f_{LO}=7.0$~GHz, output range is 10~dBm, and input range is -10~dBm. The qubit XY control signal is generated by a Z\"urich Instruments SHFSG. We use $f_{LO} = 5.4$~GHz and output range of 10~dBm.

To mitigate the microwave signal distortion caused by the electronics and control wiring, we measure the frequency response of the system through an oscilloscope. The inverse transfer function of the system is applied to the pulse samples before the samples are sent to SHFSG. This approach is significantly limited by the waveform memory of the electronics, which prohibits us to benchmark gates at longer pulse duration.

\section{DERIVATION OF AMPLIFIED LEAKAGE ERROR}
Here, we derive the amplification phases of amplified leakage error (ALE) sequence, and extract the leakage error. 

In the absence of phase and amplitude errors, the unitary of an $X$ operation with only leakage to $\state{2}$ can be written as
\begin{equation}
X = \begin{pmatrix}
0&1&-A_1e^{-i(\phi_1+\phi_2)}\\
1&0&-A_0e^{-i(\phi_0+\phi_2)}\\
A_0e^{i(\phi_0-\phi_2)}&A_1e^{i(\phi_1-\phi_2)}&e^{-i2\phi_2}
\end{pmatrix},
\label{eq:pseudo_X}
\end{equation}
where $A_0,A_1\ll1$ are the leakage population to $\state{2}$, $\phi_0,\phi_1$ are phases of the leakage population, and $2\phi_2$ is the phase acquired in $\state{2}$ during $X$. The leakage error can be calculated as $\epsilon_{\textrm{leak},\state{2}}=(A_0^2+A_1^2)/2$. A virtual-Z rotation with angle $\theta$ can be written as
\begin{equation}
\textrm{VZ}(\theta) = \begin{pmatrix}1&0&0\\0&e^{-i\theta}&0\\0&0&e^{-i2\theta}\end{pmatrix}.
\label{eq:VZ}
\end{equation}
Using Eq.~(\ref{eq:pseudo_X}) and Eq.~(\ref{eq:VZ}), every double applications of $X$ and VZ will result in, 
\begin{equation}
U=\left(\textrm{VZ}(\theta)\cdot X\right)^2= \begin{pmatrix}
1&0&u_{00}\\0&1&u_{01}\\u_{10}&u_{11}& e^{-i(4\phi_2+3\theta)}
\end{pmatrix}.
\label{eq:unitary}
\end{equation}
with $u_{ij}$ being complex elements of $U$. In order to coherently accumulate the leakage population, the amplification phases $\theta$ needs to satisfy 
\begin{equation}
4\phi_2+3\theta_k = 2k\pi. 
\label{eq:target_phases_supple}
\end{equation}
At the amplification phases $\theta_k$, the leakage population can only be amplified if it has the same phase after each operation. This indicates the ability to filter out the thermal excitation. 

To extract the leakage rate, we first obtain the effective Hamiltonian of the system,
\begin{equation}
    U^{N/2}=e^{iH_{\textrm{eff}}N/\hbar}\approx (I + i\frac{2}{\hbar}H_{\textrm{eff}})^{N/2}.
\end{equation}
Then we solve the the time dependent Schr\"odinger equation using $H_{\textrm{eff}}$,
\begin{equation}
    -\frac{i}{\hbar}H_{\textrm{eff}}\state{\Psi(N)} = \frac{d}{dN}\state{\Psi(N)},
\end{equation}
with $\state{\Psi(0)}=\state{0}$, the $\state{2}$ population after each operation can be expressed as,
\begin{align}
    P_2& = |\langle2\state{\Psi(N)}|^2= a_k\textrm{sin}^2(\omega_kN/2),
    \label{eq:population_osc}
    \\
    a_k\omega_k^2 &= A_0^2+A_1^2-2A_0A_1\textrm{cos}(\Phi+\theta_k),
    \label{eq:amplify_frequency}
\end{align}
where $N$ is the number of applications, $a_k=1/2$ for $\state{2}$, and $\Phi$ is a function of $\phi_0,\phi_1,\phi_2$. Using Eq.~(\ref{eq:amplify_frequency}) and the identity $\sum_{k=1}^n\textrm{cos}(a+2k\pi/n)=0$,
\begin{equation}
    \epsilon_{\textrm{leak},\state{2}} = \frac{1}{6}\sum_{k=1}^3 a_k\omega_k^2.
    \label{eq:leakage_extraction}
\end{equation}
In the case where $\omega_kN\ll2\pi$, the Eq.~(\ref{eq:population_osc}) reduces to $P_2 = a_k\omega_k^2N^2/4$. This can be used as the fit model when the leakage is small. 

The same analysis can be applied to $\state{3}$, and the leakage error satisfies,
\begin{align}
    P_3& = a_k\textrm{sin}^2(\omega_kN),\\
    \epsilon_{\textrm{leak},\state{3}} & = \frac{1}{10}\sum_{k=1}^5a_k\omega_k^2 = \frac{1}{2}(B_0^2+B_1^2).
\end{align}
where $B_i$ are the leakage population from $\state{i}$ to $\state{3}$. We would like to point out that the $a_k$ for $\state{3}$ depends on leakage rates from different states, including $\state{2}$, whereas the final leakage error for the $X$ operation does not. This indicates that ALE sequence is insensitive to transitions between $\state{2}$ and $\state{3}$.

\section{SINGLE QUBIT GATE CALIBRATION PROCEDURE}

\begin{figure*}[t]
    \includegraphics[width=1\linewidth]{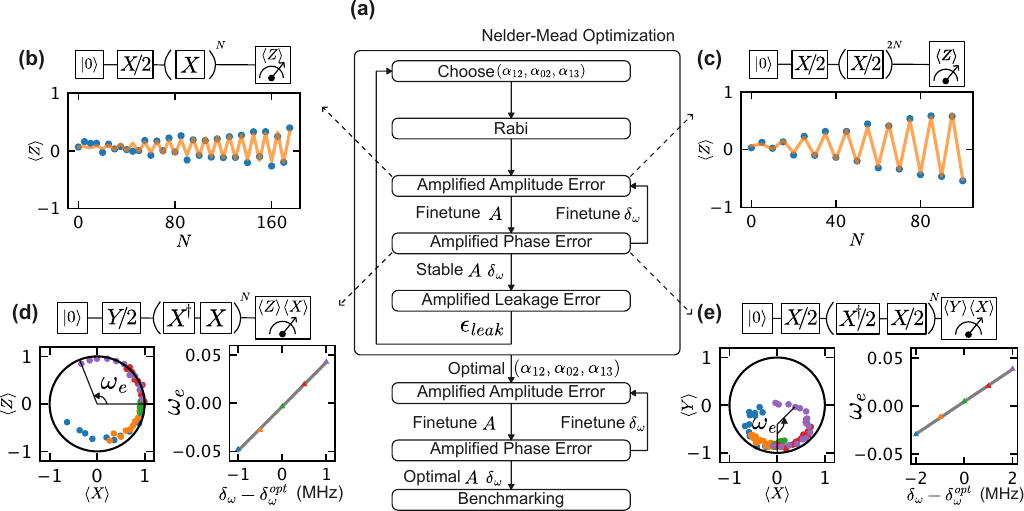}
    \caption{ Schematic of the single qubit gate calibration procedure. (a) Flow chart of the Nelder-Mead optimization. We assume the qubit frequency $f_q$ and the anharmonicity $\Delta$ is well calibrated before the Nelder-Mead optimization. (b)(c) The amplified amplitude error experiments for $X$ gate (b) and for $X/2$ gate (c). The pulse sequence, the raw result (blue dots), and the fit (orange line) are shown. (d)(e) The amplified phase error experiments for $X$ gate (d) and for $X/2$ gate (e). The pulse sequence, the raw result (colored symbols) and the fit (gray line) are shown.}
    \label{fig:calibration}
\end{figure*}

In this section, we give more details about the single qubit gate calibration procedure. We use the amplified amplitude error method (AAE) \cite{AAE}, also called ping-pong experiment, to minimize the amplitude error. We also use the amplified phase error approach (APE) \cite{APE} to minimize the phase error. In Fig.~\ref{fig:calibration}, we show the schematic of the single qubit gate calibration procedure. We assume the qubit frequency $f_q$ and anharmonicity $\Delta$ are well calibrated by Ramsey experiments before the Nelder-Mead~(N-M) optimization. In Fig.~\ref{fig:calibration}(a), we show the flow chart of the optimization procedure. For each iteration in N-M optimization, a set of $(\alpha_{12},\alpha_{02},\alpha_{13})$ is chosen by the optimizer. We first run a Rabi experiment to coarsely calibrate the pulse amplitude. Then, we use iterative AAE and APE experiments to finetune the pulse amplitude $A$ and the constant detuning $\delta_\omega$. After both parameters are stable (typically we need two iterations), an ALE experiment is performed to evaluate the leakage error $\epsilon_{leak}$. This $\epsilon_{leak}$ will be used as the loss function for optimizer to choose the next set of parameters until the algorithm converges. After the N-M optimization, extra iterations of AAE and APE are used to perform the final adjustments on the pulse amplitude $A$ and the constant detuning $\delta_\omega$.

Fig.~\ref{fig:calibration}(b)-(c) show examples of AAE experiments for both $X$ and $X/2$ gates. The results are fitted to the function
\begin{equation}
    f = a\sin\left(N(\pi+\phi_e)\right),
\end{equation}
where $\phi_e$ is the over-/under-rotation angle.

Fig.~\ref{fig:calibration}(d)-(e) show examples of APE experiments for both $X$ and $X/2$ gates. Note the differences in the initial rotation and the final measurements. The error in the constant detuning can be extracted by fitting to the model
\begin{equation}
    f = r_I\cos(\omega_eN+\phi_0)+ir_Q\sin(\omega_eN+\phi_0)+c_I+ic_Q,
\end{equation}
where the rotation frequency $\omega_e$ is proportional to the errors in the constant detuning, and $\omega_e=0$ when the constant detuning is optimal. Theoretically, the rotation radii $(r_I,r_Q)=(1,1)$ for $X$ gate and $(r_I,r_Q)=(1,0.5)$ for $X/2$ gate,  the initial angle $\phi_0 = 0$ for $X$ gate and $\phi_0 = -\pi/2$ for $X/2$ gate, the rotation centers $(c_I,c_Q) = (0,0)$ for $X$ gate and $(c_I,c_Q) = (0,-0.5)$ for $X/2$ gate.
\section{CLIFFORD-BASED RANDOMIZED BENCHMARKING}

\begin{figure}[h]
    \includegraphics[width=1\linewidth]{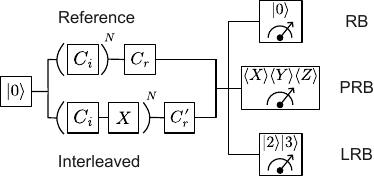}
    \caption{Schematics of all benchmarking experiments. Both reference and interleaved sequences are shown.}
    \label{fig:rb_schematic}
\end{figure}

The comprehensive benchmarking of the single qubits gates is done through standard randomized benchmarking (RB) \cite{RB}, leakage randomized benchmarking (LRB) \cite{LRB_theory}, and purity benchmarking (PRB) \cite{PRB}. We use RB to characterise the total gate error $\epsilon_{tot}$, LRB to characterize leakage error $\epsilon_{leak}$, and PRB to characterize the decoherence error $\epsilon_{decoh}$. The benchmarkings are run in an interleaved fashion to obtain the error for the target gate. See Fig.~\ref{fig:rb_schematic} for schematics of all benchmarking experiments.

All three benchmarking protocols include running multiple random sequences that consist of random Clifford gates $C_i$ in different length (N) and a final recovering gate $C_r$. 

RB measures the averaged population (P) of $\state{0}$ at the end of each sequence, and fits it to model 
\begin{equation}
    P = A+Bp^N.
    \label{eq:rb_fit}
\end{equation} 
The total gate error of the target gate operation is calculated by
\begin{equation}
    \epsilon_{tot} = (1-p_{int}/p_{ref})/2 ,
\end{equation} where $p_{ref}$ and $p_{int}$ are fitted results from reference and interleaved sequences. 

LRB measures the averaged population of leakage states, and fits it to model in Eq.~(\ref{eq:rb_fit}). 
The average leakage error of Clifford gates is $\epsilon=A(1-p)$, and the leakage error of the target gate operation is 
\begin{equation}
    \epsilon_{leak} = \epsilon_{int}-\epsilon_{ref},
\end{equation} where $\epsilon_{ref}$ and $\epsilon_{int}$ are leakage results calculated from reference and interleaved sequences. 

PRB measures the averaged purity $U = \expectation{X}^2+\expectation{Y}^2+\expectation{Z}^2$ at the end of each sequence, fits it to model
\begin{equation}
    U=A+Bu^N.
    \label{eq:prb_fit}
\end{equation} 
We use $X/2$ and $Y/2$ gate after each sequence to project the qubit state to different axes, and $\expectation{Z} = P_0-P_1$ without the normalization of $P_0+P_1=1$. The purity error of the target gate operation is calculated by 
\begin{equation}
    \epsilon_{pur}=\left(1-\sqrt{u_{ref}/u_{int}}\right)/2,
\end{equation} where $u_{ref}$ and $u_{int}$ are the fitted results from reference and interleaved sequences. As mentioned in \cite{PRB}, the decoherence error is obtained by $\epsilon_{decoh} \approx \epsilon_{pur}-\epsilon_{leak}/2$ in the presence of leakage channels.

\begin{figure}
    \includegraphics[width=1\linewidth]{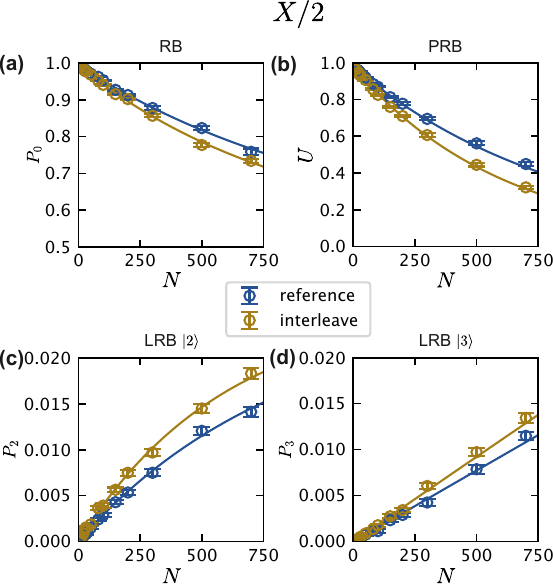}
    \caption{ Benchmarkings of the optimal $X/2$ gate at $t_p=6.8$~ns, averaged over 60 random sequences. The error bar indicates the standard error of the mean. The solid lines are fit curves. (a) RB. (b) PRB. (c) LRB for $\state{2}$. (d) LRB for $\state{3}$. }
    \label{fig:rb_sx}
\end{figure}

\begin{figure}
    \includegraphics[width=1\linewidth]{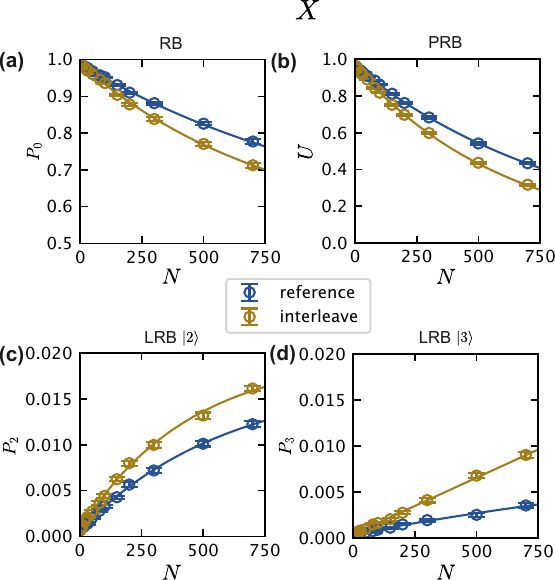}
    \caption{Benchmarkings of the optimal $X$ gate at $t_p=6.8$~ns, averaged over 60 random sequences. The error bar indicates the standard error of the mean. The solid lines are fit curves. (a) RB. (b) PRB. (c) LRB for $\state{2}$. (d) LRB for $\state{3}$.}
    \label{fig:rb_x}
\end{figure}
For the Clifford gates decomposition, we follow the mapping stated in Ref.~\cite{clifford_decomposition}. The $t_p=8$~ns $X$ gate is used when the pulse lengths of $X/2$ gate are swept, and vice versa. As the waveform memory of the control electronics is limited, the benchmarkings cannot reach the saturation values. We use $A = (\Lambda^{ps}_{01}+\Lambda^{ps}_{10})/2$ for RB fitting, where $\Lambda^{ps}_{ij}$ are the elements in the assignment matrix. We use $A=0$ for PRB fitting.  In the case of low leakage population in LRB, we use $P=(1-p)AN$ as the fit model and assume $A=-B$. Fig.~\ref{fig:rb_sx} and Fig.~\ref{fig:rb_x}, we show the benchmark curves for the optimal $X/2$ and $X$ gates at $t_p=6.8$~ns, respectively. All benchmark curves are averaged over 60 random sequences.

\begin{figure}
    \includegraphics{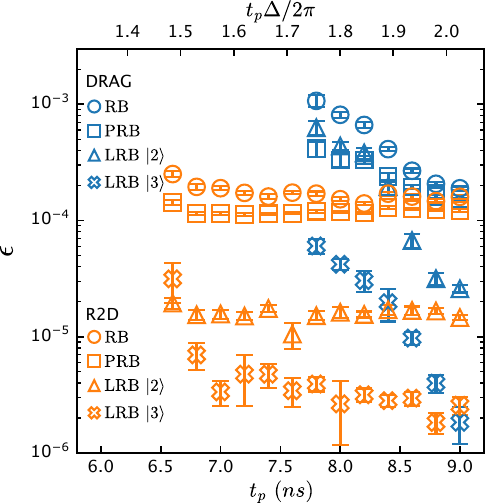}
    \caption{Benchmarking results for $X$ gates using the standard DRAG and the R2D methods. R2D outperforms standard DRAG.}
    \label{fig:DRAG_rbs}
\end{figure}

We directly compare the benchmarking results for $X$ gates using standard DRAG and R2D corrections in Fig.~\ref{fig:DRAG_rbs}. For the standard DRAG correction, the $\alpha_{12}$ is optimized to minimize the total leakage error, and $\alpha_{02}$ and $\alpha_{13}$ are set to 0. 
\clearpage
\end{document}